\documentclass[aps,prd,reprint,superscriptaddress]{revtex4-1}

\usepackage[T1]{fontenc}

\usepackage[english]{babel} 
\usepackage{wasysym}
\usepackage{graphicx}
\usepackage{shadow}
\usepackage{epsfig}
\usepackage{natbib}
\usepackage{dcolumn}
\usepackage{hyperref}
\usepackage{float}
\hypersetup{
	pdfstartview=FitH,
	pdfpagemode=UseNone,
hidelinks
}
\newcommand{\CP}{$CP$}

\newcommand{\eqref}[1]{(\ref{#1})}

\begin{document}


\title{Constraining short-range spin-dependent forces with polarized $^3$He}

\author         {M. Guigue}
\email          {guigue@lpsc.in2p3.fr}
\affiliation{LPSC, Universit\'e Grenoble-Alpes, CNRS/IN2P3, Grenoble, France}
\author{D. Jullien}
\affiliation{Institut Laue Langevin, 53, Rue Horowitz, 38000 Grenoble}
\author{A. K. Petukhov}
\affiliation{Institut Laue Langevin, 53, Rue Horowitz, 38000 Grenoble}
\author         {G. Pignol}
\email          {pignol@lpsc.in2p3.fr}
\affiliation{LPSC, Universit\'e Grenoble-Alpes, CNRS/IN2P3, Grenoble, France}

\date{\today}

\begin{abstract}
We have searched for a short-range spin-dependent interaction using the spin relaxation of hyperpolarized $^3$He. 
Such a new interaction would be mediated by a hypothetical light scalar boson with \CP-violating couplings to the neutron.
The walls of the $^3$He cell would generate a pseudomagnetic field and induce an extra depolarization channel.
We did not see any anomalous spin relaxation and we report the limit for interaction ranges $\lambda$ between $1$ and $100~\rm{\mu m}$: $g_sg_p \lambda ^2 \leq 2.6\times 10^{-28}~\mathrm{m^2}\,  ( 95~\%\,  \mathrm{C.L.})$, where $g_s$($g_p$) are the (pseudo)scalar coupling constant, improving the previous best limit by 1 order of magnitude.
\end{abstract}
\maketitle

\section{Introduction}
Theories beyond the Standard Model (SM) of particle physics generically predict the existence of new particles.
They can broadly be divided into two categories, \emph{ultraviolet} and \emph{infrared} modifications of the SM. 

Concerning the ultraviolet category, new particles are expected with masses above the electroweak scale of about $100$~GeV. 
For example, grand unified theories are associated with an energy scale as high as $10^{15}~\rm{GeV}$. 
In the case of supersymmetry and similar theories, new particles could be discovered just above the electroweak scale. 
These particles are still actively searched for in high energy proton-proton collisions at the Large Hadron Collider. 
Although it is perhaps too early to reach a definitive conclusion, as a matter of fact no evidence for the existence of new physics in the TeV range has been reported by Run 1 of the LHC. 

Alternatively, new physics could manifest itself in the infrared, that is, at energies much below the electroweak scale (see Ref. \citep{Jaeckel2010} for a review on the low energy frontier of particle physics). 
In this case, one should look for weakly interacting slim particles (WISPs), having masses below $1~\rm{eV}$. 

Nambu--Goldstone scalar bosons, arising from the spontaneous breaking of a global symmetry, establish a well-motivated theoretical case for WISPs. 
These bosons are naturally predicted to be very light, even massless if the symmetry is not explicitly broken. 
The prominent example is the hypothetical QCD axion, the boson associated with the global $U(1)$ Peccei--Quinn symmetry introduced to solve the strong \CP problem \citep{Kim2010}.
The QCD axion is basically a one-parameter theory: its mass and couplings to SM particles can be derived from a single symmetry breaking scale parameter $f_a$, which lies presumably between $10^{9}$ and $10^{12}~\rm{GeV}$. 
One can imagine other Nambu--Goldstone bosons with no specific relation between the mass, symmetry breaking scale and the couplings to SM particles, these bosons being referred to as axionlike particles or ALPs. 
It is interesting to note that although ALPs induce new phenomena in the infrared (because they are very light), they are associated with a symmetry breaking at an energy scale well above the electroweak scale, as it is the case for the QCD axion. 
Other than spin-0 ALPs, there can exist new light spin-1 bosons, arising from  broken hidden $U(1)$ symmetries, which do no decouple from SM particles even in the limit of vanishing mass \citep{Fayet1990}.  

The existence of dark matter provides another motivation to search for new light particles, since it turns out that WISPs are suitable dark matter candidates \citep{Ringwald2012a,Arias2012a}.
Experiments aiming at detecting WISPy dark matter, called \emph{haloscopes}, are very different from detectors searching for weakly interacting massive particles. 
Instead of catching hits of single particles, one needs to detect an oscillating field. 
For example, ADMX \citep{Asztalos2010} uses a resonant cavity to convert these oscillations to microwave photons, provided the WISP couples with the photon. 

Regardless of the WISPs being the dark matter, astrophysical objects like the Sun or supernovae could be a sizable source of WISPs via thermal production. 
A too big escape channel for WISPs would contradict the observed time scale of the evolution of stars. 
This "energy loss" argument provides stringent bounds on the coupling of WISPs with SM particles \citep{Raffelt1995}. 
Experiments such as CAST \citep{Arik2014} aim at detecting the flux of WISPs emitted by the Sun, by converting them into x-ray photons in a so-called \emph{helioscope}. 

It is desirable to design laboratory experiments to search for WISPs that do not rely on a cosmological or astrophysical source. 
Only the combination of independent results can confirm or discard the existence of exotic bosons.
One type of laboratory experiments is \emph{light shining through a wall} (see Ref. \citep{Redondo2011} for a review) probing the WISP-photon coupling.
Another type of experiments, to probe the couplings to fermions, consists of searching for a \textit{fifth force}. 
The exchange of an exotic light boson between two fermions induces a macroscopic force which is derived from a Yukawa potential \cite{Moody1984}
\begin{equation}
V(x) =  \frac{g_s^2}{4\pi}\frac{\hbar c}{x}\exp (-x/\lambda ),
\label{eq:Yukawa}
\end{equation}
with $g_s$ the coupling constants of the boson to the fermions at the interaction vertex and $x$ the distance between the two fermions.
The interaction range $\lambda$ is given by
\begin{equation}
\lambda = \frac{\hbar c}{m_0 c^2},
\label{eq:lambda}
\end{equation}
with $m_0$ the mass of the light boson.
While the range of the interaction induced by a heavy boson like the $W$  ($m_W = 80~\rm{GeV}$) is as short as $10^{-15}~\rm{m}$, a boson lighter than $0.1~\rm{eV}$ will generate macroscopic effects between objects separated by $2~\rm{\mu m}$.
Numerous experiments searching for this fifth force have been realized for a wide range of boson masses (see Ref. \citep{Antoniadis2011} for a review of this topic).

The potential \eqref{eq:Yukawa} corresponds to the most simple version with a scalar coupling $g_s$ to fermion, resulting in a monopole-monopole interaction. 
Now, in the presence of a pseudoscalar coupling $ig_p \gamma ^5$, the interaction becomes \emph{spin dependent} and cannot be discovered by fifth force experiments using macroscopic bodies.
In this paper, we report on a search for a spin-dependent interaction of the type
\begin{equation}
V(x) = g_{s}g_{p} \frac{n\lambda\hbar^2}{4m}\sigma _z e^{-x/\lambda},
\label{eq:potential}
\end{equation}
generated by a macroscopic source of unpolarized nucleons with volumic density $n$ acting on a polarized probe with spin $\sigma_z$ and mass $m$.
The nucleons occupy an infinite thick plate and the probe is situated outside of the plate, at a distance $x$ from the surface. 
The constant $g_s$ is the scalar coupling of the source nucleon to the boson, and $g_p$ is the pseudoscalar coupling constant between the probe and the boson.
The potential \eqref{eq:potential} corresponds to the sum of two monopole-dipole potentials $\mathcal{V}_{9,10}$ presented in Dobrescu and Mocioiu 's classification \citep{Dobrescu2006}.

As a practical realization, we considered a $^3$He polarized gas contained in a glass cell.
The source of the interaction acting on the $^3$He spins is the nucleons in the walls of the cell.
The helium spins then probe a short-range pseudomagnetic field 
\begin{equation}
b(x) = g_{s}g_{p} \frac{\lambda\hbar}{2m\gamma}e^{-x/\lambda},
\label{eq:champ-pseudo-total}
\end{equation}
with $\gamma /2\pi = 32.4~\rm{Hz/\mu T}$ \citep{Flowers1993} the gyromagnetic ratio of the $^3$He atoms.
The motion of the spins in this pseudomagnetic field will induce an anomalous longitudinal depolarization of the gas, in addition to the usual depolarization mechanisms.
By studying the longitudinal relaxation rate of the $^3$He gas as a function of the applied magnetic field, this effect can be separated from the other standard contributions.

In the case of the QCD axion, when taking into account the bound on the QCD \CP-violating phase $\theta <10^{-10}$ derived from the most recent measurement of the neutron electric dipole moment (nEDM)\cite{Pendlebury2015}, the product of the coupling constants $g_sg_p$ is predicted to be typically smaller than $10^{-26}$ for a $1~\rm{\mu m}$ range \cite{Moody1984}, which is more than 10 orders of magnitude below the current sensitivity.
However, in the generic case of ALPs, no such prediction exists.

In 2010, a preliminary experiment \citep{Petukhov2010} performed at the Institut Laue-Langevin (ILL) measured  the depolarization rate of a hyperpolarized $^3$He gas and set a competitive constraint on the $g_sg_p$ coupling.
To improve our sensitivity, we built a dedicated setup at the ILL. 
The enhancement of sensitivity, as compared to other techniques, comes from the very long relaxation time (several days under certain conditions) of the polarized gas. 

In Sec. II, we will review the theory of depolarization of particles moving in an inhomogeneous field.
The expression of the anomalous depolarization induced by the pseudomagnetic field will be also derived.
The experimental apparatus will be presented in details in Sec. III.
Finally, the results of data analysis and the obtained constraints will be shown in Sec. IV.

\section{Theory of standard and exotic spin relaxation of $^3$He}

We consider the case of an assembly of spin-1/2 particles with a gyromagnetic ratio $\gamma$ contained in a glass cell of volume $V$ and immersed in a holding magnetic field $\vec{B}_0 = B_0 \vec{e}_z$.
The magnetic inhomogeneities are quantified by $\vec{b} = \left( b_x, b_y, b_z\right)$.
The Larmor precession frequency of the spins is $\omega _0 = \gamma B_0$.
The cell walls of thickness $d$ act as a source of a pseudomagnetic field felt by a polarized particle
\begin{equation}
\vec{b}_a(x) = b_a e^{-x/\lambda} \vec{e}_x ,
\label{eq:shape-pseudofield}
\end{equation}
orthogonal to the wall surface with $b_a = g_{s}g_{p} \frac{n\lambda\hbar}{2m\gamma}\left( 1-e^{-d/\lambda}\right)$.
The distance between the wall and the polarized particle is given by $x$.
The density of nucleons in the glass (typically $1.6\times 10^{30}~\mathrm{nucleons}/\mathrm{m}^3$) is denoted $n$.
The field components which are transverse to the holding field direction will induce a longitudinal relaxation of the particles polarization.
The contribution to the relaxation induced by this short-range pseudomagnetic field \eqref{eq:shape-pseudofield} has a very peticular dependence on the holding magnetic field $B_0$: in certain conditions, the relaxation rate is proportional to $1/\sqrt{B_0}$, as we will show later on.
In fact, the other depolarization channels one can expect behave as powers of $1/B_0$.
In this section, we will discuss the expected contributions to the depolarization.
We will then use Redfield's theory of relaxation to calculate the relaxation rate associated with the magnetic inhomogeneities and by the sought short-range pseudomagnetic field. 

\subsection{Standard sources of relaxation}

Since the $^3$He gas polarization $P$ is much larger than the polarization at the thermal equilibrium ($P(t=0)\approx 70~\%$), the gas will depolarize as
\begin{equation}
P(t) = P(t=0)\exp (-\Gamma _1 t),
\end{equation}
with $\Gamma _1$ the longitudinal relaxation rate.
Three main phenomena contribute to the relaxation: : collisions with the cell walls, collisions between $^3$He atoms and particles motion in an inhomogeneous magnetic field.
It is commonly known that the first contribution, quantified by $\Gamma _w$, does not depend on the gas polarization, its pressure and the holding field value \citep{Jacob2001,Saam2012}.
The second contribution $\Gamma _{dd}$ depends on the frequency of atomic collisions and thus is proportional to the pressure of the gas \citep{Newbury1993}.
The last contribution is denoted $\Gamma _m$ and corresponds to a perturbation of spins induced by their motion in an inhomogeneous magnetic field.
Each atom effectively sees a fluctuating magnetic field, of which transverse components fluctuations at the Larmor frequency induce spins inversions.
This contribution has been discussed for decades in the literature. 
The calculation of the corresponding rate is presented in the next section.

\subsection{Redfield theory and standard magnetic relaxation}

To express the spin relaxation rate $\Gamma _m$ of a polarized gas in slightly inhomogeneous magnetic fields, the Redfield theory \citep{Redfield1965} can be applied when the resulting relaxation time $T_1$ is much longer than the decay time of the magnetic field correlation functions.
This condition is satisfied for a large variety of systems, such as a $^3$He polarized gas at atmospheric pressure and immersed in a several $\rm{\mu T}$ holding field.

The relaxation rate of the gas can then be expressed as the Fourier transform of the transverse components correlation functions at the Larmor frequency of the spins
\begin{equation}\label{eq:long-relaxation-rate}
\Gamma _m=\frac{1}{T_1}=\gamma ^2  \int _0^{\infty} \langle b_x (0)b_x (\tau) + b_y (0)b_y (\tau)\rangle  \cos (i\omega \tau ) \mathrm{d}\tau .
\end{equation}
The ensemble average (over the articles in the cell) is denoted $\langle \cdots\rangle$.
The correlation function of the magnetic components $b_i$ and $b_j$ is expressed as
\begin{equation}
\langle b_i(0)b_j(\tau )\rangle = \frac{1}{V} \int _V \mathrm{d}\vec{r}_0\int _V \mathrm{d}\vec{r} b_i(\vec{r}_0)b_j(\vec{r}) \pi (\vec{r},\tau\vert\vec{r}_0) .
\label{eq:corrfunction}
\end{equation}

The function $\pi (\vec{r},\tau\vert\vec{r}_0)$ corresponds to the conditional probability (also called the propagator) for a particle, being at $\tau = 0$ at $\vec{r}_0$, to be at $\tau$ at $\vec{r}$.
In a general way, $\pi (\vec{r},\tau\vert\vec{r}_0)$ satisfies the initial condition
\begin{equation}
\pi (\vec{r},\tau =0 \vert\vec{r}_0) = \delta (\vec{r}-\vec{r}_0) .
\end{equation}
and the boundary condition
\begin{equation}
\vec{\nabla}\pi (\vec{r},\tau|\vec{r}_{0})\cdot \vec{n}=0
\label{eq:boundaries} .%
\end{equation}
In the case of gases at several bars, the gas is in "diffusive regime" and the propagator is governed by the diffusion equation
\begin{equation}
\frac{\partial \pi }{\partial \tau} = D\Delta \pi ,
\label{eq:diffusion}
\end{equation} 
with $D$ the diffusion coefficient of the gas.
In the case of $^3$He gas at a pressure of $1~\rm{bar}$, $D\approx 1.84~\rm{cm^2/s}$ \citep{Barbe1974a,Hayden2004}.

For  gases in the diffusive regime immersed in holding magnetic fields of several $\rm{\mu T}$, there are several sources of magnetic inhomogeneities.
For a given cell filled with a helium gas at a pressure $p$ and a polarization $P$, the magnetic inhomogeneities are the sum of three terms:
\begin{equation}
\vec{b} = \vec{b}_{\mathrm{ext}} + \vec{b}_{\mathrm{coil}} (B_0) + \vec{b}_{\mathrm{cell}} (P,p) .
\label{eq:realmagfield}
\end{equation} 
The term $\vec{b}_{\mathrm{ext}}$ corresponds to magnetic inhomogeneities induced by the apparatus environment, and $\vec{b}_{\mathrm{coil}} (B_0)$ corresponds to the one created by the holding field generator and so is proportional to $B_0$.
The last contribution $\vec{b}_{\mathrm{cell}} (P,p)$ is due to the fact that at high pressure and polarization, the $^3$He gas behaves as a magnet and generates an inhomogeneous magnetic field proportional to $p$ and $P$.

For an arbitrary geometry cell immersed in a magnetic field with an arbitrary spatial profile, the relaxation rate can be estimated \citep{Guigue2014} with the relation 
\begin{equation}
\Gamma _m = D\frac{\langle (\vec{\nabla} b_{\perp})^2\rangle}{B_0^2},
\label{eq:adiabdiff}
\end{equation}
where $\langle (\vec{\nabla} b_{\perp})^2\rangle$ corresponds to the average over the cell volume of the squared transverse gradients. 
Using \eqref{eq:realmagfield}, we can write the total longitudinal relaxation rate for a given pressure as
\begin{equation}
\Gamma _1 =  a + \frac{b}{B_0} + \frac{c}{B_0^2} + \frac{dP}{B_0} + \frac{eP}{B_0^2} + \frac{fP^2}{B_0^2} ,
\label{eq:Gamma1-explicite}
\end{equation}
with $a$, $b$, $c$, $d$, $e$ and $f$ real coefficients.
For a given pressure in a given cell, the parameter $a$ in \eqref{eq:Gamma1-explicite} corresponds to the sum of the relaxation rates induced by atomic $\Gamma _{dd}$ and walls $\Gamma _w$ collisions, the spin-flip induced relaxation and the magnetic depolarization generated by the solenoid gradients. 
The third term $c$ in \eqref{eq:Gamma1-explicite} corresponds to the relaxation induced by the environmental magnetic inhomogeneities $\vec{b}_{\mathrm{ext}}$. 
The last term $f$ in this equation corresponds to the depolarization induced by the field gradient generated by the polarized gas. 
The other coefficients $b$, $d$ and $e$ correspond to interference terms of magnetic gradients which have very different origins.
For example, the coefficient $b$ corresponds to the average value of the product of the solenoid magnetic gradients (which therefore depends on $B_0$) and the magnetic gradients induced by the environment (which does not depend on $B_0$). 
Coefficients $a$, $c$ and $f$ which have a clear physical meaning will be discussed in details in Sec. \ref{sec:bahviorfitparam}. 

\subsection{Relaxation rate induced by a short-range spin-dependent interaction}

If one assumes an exotic interaction \eqref{eq:potential} between the $^3$He spins and the cell walls induced by a light boson, a new depolarization channel will be added to the standard relaxation \eqref{eq:Gamma1-explicite}.
For simplicity, let us consider a one-dimensional problem: the source of the short-range pseudomagnetic field \eqref{eq:shape-pseudofield} is a glass plane of surface $S$ and thickness $d$, placed in $x=0$.
The polarized particles evolve at the right side of the source for $x$ positive.
In this case, the constant $b_a$ corresponds to the amplitude of the pseudomagnetic field (\ref{eq:shape-pseudofield}) generated by the wall felt by the probe particle,
\begin{equation}
b_a = g_sg_p\frac{\hbar n\lambda}{2m \gamma} \left( 1-e^{-d/\lambda}\right) ,
\end{equation}
with $n$ the density of nucleons in the glass (typically $1.6\times 10^{30}~\mathrm{nucleons}/\mathrm{m}^3$).
In the case of a polarized gas at atmospheric pressure and magnetic inhomogeneities larger than the mean free path of the probe particles, the diffusion equation (\ref{eq:diffusion}) describes correctly the particle motion.
The particle propagator can be expressed as a continuous sum of cosines,
\begin{equation}
\pi (\vec{r},\tau \vert \vec{r}_0) = \frac{1}{S\pi} \int _{-\infty} ^{\infty} dk \cos (kx) \cos (kx_0) \exp (-\tau Dk^2) . 
\end{equation}
This expression assumes that the system is invariant under translations along $\vec{e}_y$ and $\vec{e}_z$.
One can calculate the correlation function $\langle b (0) b(\tau)\rangle$ with $b$ the pseudomagnetic component orthogonal to the surface,
\begin{equation}
\resizebox{1.\hsize}{!}{$\displaystyle{\langle b(0)b(\tau )\rangle = \frac{b_a^2}{V\pi S}\int _{-\infty} ^{\infty} \mathrm{d}k \left( \int _S \mathrm{d}y \mathrm{d}z \int _0 ^{L} \mathrm{d}x e^{ -x /\lambda} \cos (kx)\right) ^2 e^{ -\tau Dk^2 }}$},
\end{equation}
where $L$ is defined as the  ratio $V/S$.
The length $L$ corresponds to  the characteristic distance between the walls of the cell.
If the interaction range $\lambda$ is much smaller than $L$, we can write 
\begin{equation}
\langle b(0)b(\tau )\rangle = \frac{b_a^2S}{V\pi}\int _{-\infty} ^{\infty} \mathrm{d}k \left( \frac{\lambda }{1+k^2\lambda ^2}\right) ^2 e^{ -\tau Dk^2 }.
\label{eq:bb}
\end{equation}
The relaxation rate (\ref{eq:long-relaxation-rate}) induced by a wall acting as a pseudomagnetic field source on a polarized particle is the Fourier transform of the correlation function $\langle b(0)b(\tau )\rangle$ (\ref{eq:bb}) at the frequency $\omega _0$
\begin{eqnarray}
& & \resizebox{1.\hsize}{!}{$\displaystyle{\Gamma _{NF} = \frac{\gamma ^2}{2} \frac{b_a^2S}{V\pi} \int _0 ^{\infty} \mathrm{d}\tau \cos (\omega _0 \tau )\int _{-\infty} ^{\infty} \mathrm{d}k \left( \frac{\lambda }{1+k^2\lambda ^2}\right) ^2 e^{ -\tau Dk^2 }}$}\nonumber\\
&=& \left( \gamma b_a\right) ^2 \frac{S}{V} \frac{\lambda ^3}{2D}\frac{1}{(1+\phi _{\lambda}^2)^2}\nonumber\\
&\times& \left( \sqrt{\frac{2}{\phi _{\lambda}}}\left( 1-\phi _{\lambda}\left( \phi_{\lambda}-2\right)\right) +\phi_{\lambda}^2-3\right) ,
\label{eq:Gamma1NF-1plaqueinfinie}
\end{eqnarray}
with $\phi _{\lambda} = \omega _0\frac{\lambda ^2}{D}$.
The general solution of the one-dimension case of a polarized gas in the diffusive regime contained between two plates  was treated in Ref. \citep{Petukhov2010}, using the Redfield theory and a propagator expansion with sines and cosines.
Our result \eqref{eq:Gamma1NF-1plaqueinfinie} is only valid in the limit $\omega _0 L^2/D \gg 1$.
 
In practice, polarized gases are contained in cubic, cylindrical or spherical cells with a wall thickness $d$.
The longitudinal relaxation rate \eqref{eq:Gamma1NF-1plaqueinfinie} can be adapted for any shape of cell with a typical size $L=V/S$, much larger than the interaction range $\lambda$.
With this assumption, one can assume that the walls surface is flat: the relaxation rate induced by a short-range field for a cell is proportional to the calculated rate \eqref{eq:Gamma1NF-1plaqueinfinie}, valid for a flat surface.
We can define the \emph{apparent surface} $S_a$, corresponding to the cell surface which effectively contributes to the short-range relaxation
\begin{equation}
S_a = \int _S \mathrm{d}S \left[ \left( \vec{e}_x\cdot \vec{n}\right) ^2 + \left( \vec{e}_y\cdot \vec{n}\right) ^2\right] ,
\label{eq:appsurf}
\end{equation}
with $\vec{n}$ the unitary vector orthogonal to the surface.
For example, the apparent surface of a sphere with radius $R$ is $8\pi R^2/3$.
Finally, the longitudinal relaxation rate of a polarized gas contained in a cell with a volume $V$ and a typical size $L$ larger than $\lambda$ can be written, using the apparent surface $S_a$, as
\begin{eqnarray}
\nonumber
\Gamma _{NF} &=& \left( \gamma b_a\right) ^2 \frac{S_a}{V} \frac{\lambda ^3}{2D}\frac{1}{(1+\phi _{\lambda}^2)^2}\\
&\times& \left( \sqrt{\frac{2}{\phi _{\lambda}}}\left( 1-\phi _{\lambda}\left( \phi_{\lambda}-2\right)\right) +\phi_{\lambda}^2-3\right) .
\label{eq:Gamma1NF-formulefinale}
\end{eqnarray}
In the case of a scalar-pseudoscalar interaction between a $^3$He polarized gas and the nucleons in the cell walls, the relaxation rate is thus proportional to $\left( g_sg_p\right) ^2$.
Moreover, for $\phi _{\lambda} \ll 1 \ll \omega _0 L^2/D$ , Eq. \eqref{eq:Gamma1NF-formulefinale} simplifies into
\begin{equation}
\Gamma _{NF} = \left( \gamma b_a\right) ^2 \frac{S_a}{2V}\sqrt{\frac{2\lambda ^2}{D\omega _0}},
\label{eq:GammaNF_simple}
\end{equation}
which is valid for any shape of cell.
In this regime, the behavior of the relaxation rate with respect to the holding field is very different from what could be expected \eqref{eq:Gamma1-explicite}.
From Eq. \eqref{eq:GammaNF_simple} immediately follows that the behavior of the relaxations due to short-range forces in spherical and cubic cells are exactly the same since the factor $S_a/V$ is equal to $2/R$ for a sphere with a radius $R$ and $4/s$ for a cube with $s$ the length of its edge. 

The search of an exotic short-range interaction can be performed using polarized $^3$He by measuring the relaxation of the gas as a function of the holding field and the polarization.
If a deviation of the behavior of this rate is observed and corresponds to the induced short-range contribution \eqref{eq:Gamma1NF-formulefinale}, the existence of a new boson will be revealed.

\section{Experimental setup and procedure}

The experimental method is rather straightforward in principle, it simply consists in measuring the relaxation rate of $^3$He cells for different values of the holding magnetic field (and also for different values of the polarization of the cell). 
To polarize cells we have used the "Tyrex" Metastability Exchange Optical Pumping (MEOP) installation, described in details elsewhere \citep{Andersen2005}. 
The installation can provide up to 4 bars of polarized $^3$He in a valved glass cell, at an initial polarization of up to $75~\%$. 
The cell is then transferred to the apparatus dedicated  to the measurement of the decay of the polarization, situated in the same building at the ILL. 
This apparatus is designed to operate a large range of possible holding magnetic fields $B_0$ ranging from $3$ to $300~\rm{\mu T}$. 
In this section we present in detail the setup. 

\subsection{Magnetic environment}

To decrease the magnetic depolarization contributions, the measurement of the gas polarization is performed in a very well-controlled magnetic environment.
This apparatus is composed of a cylindrical mumetal magnetic shield and a solenoid.

A part of the mumetal tube, formerly used for the neutron-antineutron oscillations experiment at the ILL \citep{Baldo-Ceolin1994}, was refurbished.
This large tube \citep{Bitter1991a} ($1~\rm{m}$ diameter, $4.5~\rm{m}$ long and $0.8~\rm{mm}$ thick) acts as a magnetic screen to shield the ambient magnetic field.

\begin{figure}
	\centering
		\includegraphics[width=0.470\textwidth]{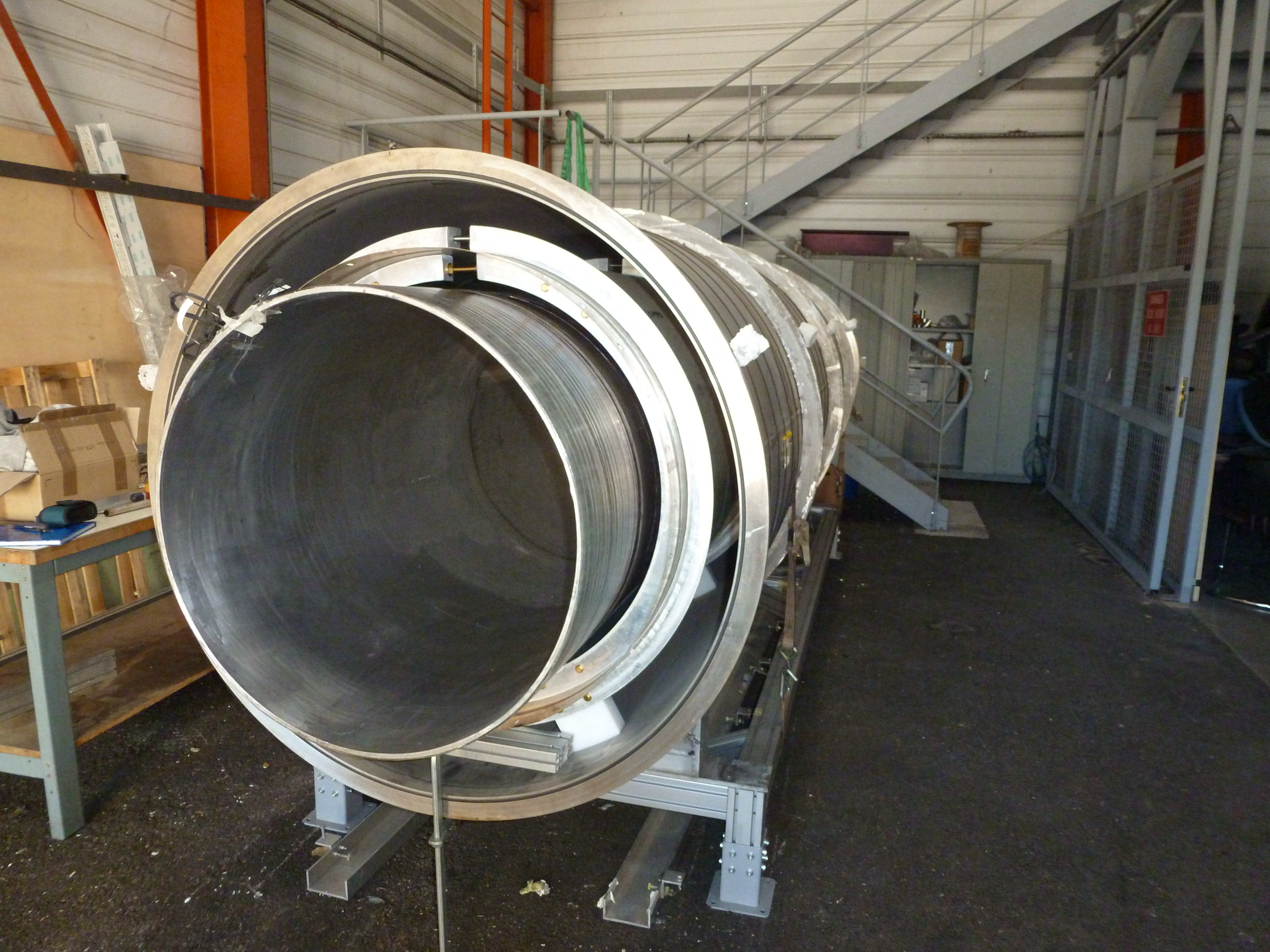}
	\caption{Solenoid inserted into the mumetal magnetic shield.}
	\label{fig:vue_ensemble}
\end{figure}

The $5~\rm{m}$ long and $80~\rm{cm}$ diameter solenoid is composed of 2363 spires around a $5~\rm{mm}$ aluminum tube.
It was installed inside the magnetic shield to provide a tunable, stable and homogeneous magnetic field in a large enough volume.
Figure \ref{fig:vue_ensemble} shows the solenoid and the magnetic shield.
For $1~\rm{A}$ circulating current, the longitudinal magnetic field generated in the center of the solenoid is $590~\rm{\mu T}$.

We performed several field maps (in a volume of $20\times 20\times 20~\rm{cm^ 3}$) using a three-axis fluxgate magnetometer for various values of the holding field $B_0$. 
The maps were analyzed with a second-order 3D polynomial fit to extract the value of the transverse gradients $g_{\perp}$.
The results are presented in Table \ref{tab:RésultatsDesCartographiesDeChampsMagnétiques}: typically, $g_{\perp}$ is about $2.4$ to $4.2\, \rm{nT/cm}$ for magnetic field from $2$ to $80\, \rm{\mu T}$.
\begin{table}
	\centering
	\caption{Results of the magnetic field mapping.
	The relaxation rates are calculated for a spherical $5~\rm{cm}$ radius cell filled with $1$ bar of $^3$He.}
	\label{tab:RésultatsDesCartographiesDeChampsMagnétiques}
		\begin{tabular}{c|c|c}
			$B_0$ [$\mu$T] & $\sqrt{g_{\perp}^2}$ [nT/cm] & $\Gamma _{1m,\rm{calculated}}$ [h$^{-1}$] \\
			\hline
			2.05  &  2.42 & $9.2\times 10 ^{-3}$ \\
			10.47 &  2.47 & $3.7\times 10 ^{-4}$ \\
			21.20 &  2.63 & $1.0\times 10 ^{-4}$ \\
			83.81 &  4.21 & $1.7\times 10 ^{-5}$
		\end{tabular}
\end{table}
These weak magnetic gradients will give very long relaxation times $T_1$: a calculation from Eq. \eqref{eq:adiabdiff} gives times longer than $110\, \rm{h}$ for magnetic fields higher than $2~\rm{\mu T}$.
Our magnetic apparatus can therefore be used to maintain $^3$He cells polarized for several days.

\subsection{Direct polarimetry}

The spin relaxation rate is extracted by periodically measuring the polarization of a $^3$He cell. 
We chose to use a direct polarimetry technique, first presented by Cohen Tannoudji \citep{Cohen-Tannoudji1969}. 
It consists of recording the magnetic field generated by the magnetized cell itself, which is proportional to the polarization $P$. 
This is possible using commercial magnetometers since a fully polarized cell filled with $1\,\rm{atm}$ of $^3$He generates a dipolar field of tens of nT. 

\begin{figure}
\includegraphics[width=0.320\textwidth]{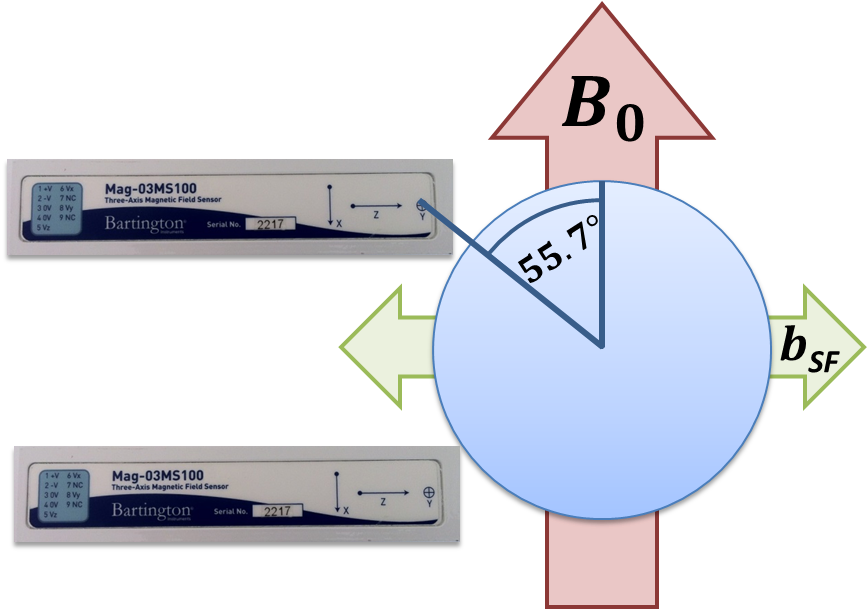}
\caption{\label{fig:SF-apparatus} Scheme of the spin-flipping and measurement apparatus.
The spherical cell is positioned in contact with the two magnetometers inside the spin-flip coil which generates an oscillating magnetic field $b_{SF}$ transversely to the holding magnetic field $B_0$.}
\end{figure}

\begin{figure}
	\centering
		\includegraphics[width=0.470\textwidth]{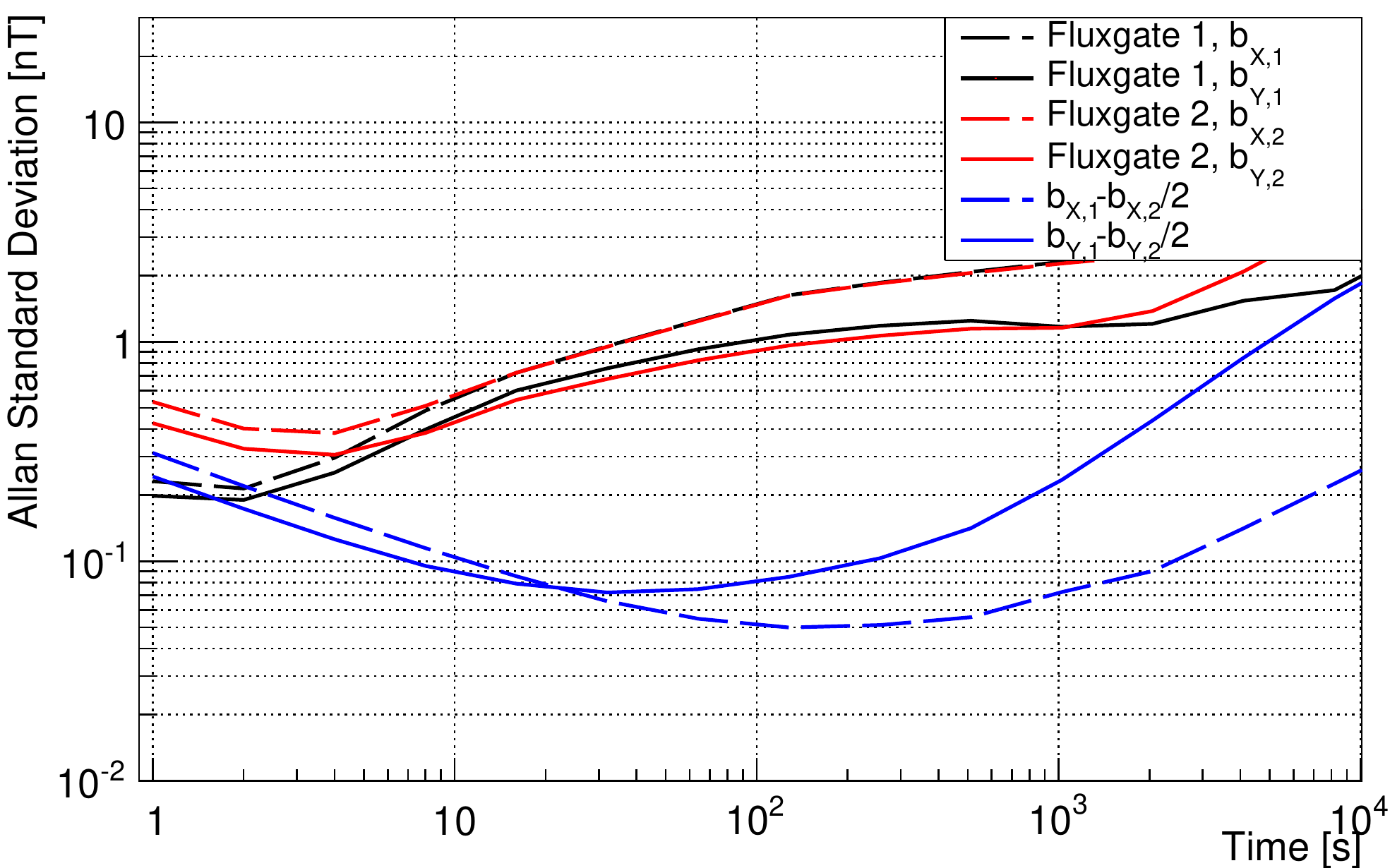}
	\caption{Allan Standard Deviation (ASD) of the magnetic components transverse to the holding field direction for $B_0 = 80~\rm{\mu T}$. 
	The dashed lines correspond to the $X$ component and the continuous lines to the $Y$ component ($X$ and $Y$ are two directions orthogonal to $B_0$). 
	In black, the ASD of fluxgate 1; in red, the ASD of fluxgate 2; in blue, the difference between the two fluxgates divided by 2. }
	\label{fig:ASD_Stab_Btrans}
\end{figure}

Two Bartington low noise fluxgate magnetometers are placed where the dipolar field induced by the polarized gas is transverse relative to the $B_0$ field, as represented on Figure \ref{fig:SF-apparatus}.
This configuration with two fluxgates allows one to compensate for the random fluctuations of the environmental transverse fields by taking the difference between the two magnetometer readings, as the dipolar fields created by the polarized $^3$He gas at positions 1 and 2 are opposite. 
The performances of the setup in terms of the time stability are quantified by the Allan Standard Deviation shown in Figure \ref{fig:ASD_Stab_Btrans} for a holding field value of $80~\rm{\mu T}$: the difference reduces the long-time correlated fluctuations of the two fluxgates.

\begin{figure}
	\centering
		\includegraphics[width=0.47\textwidth]{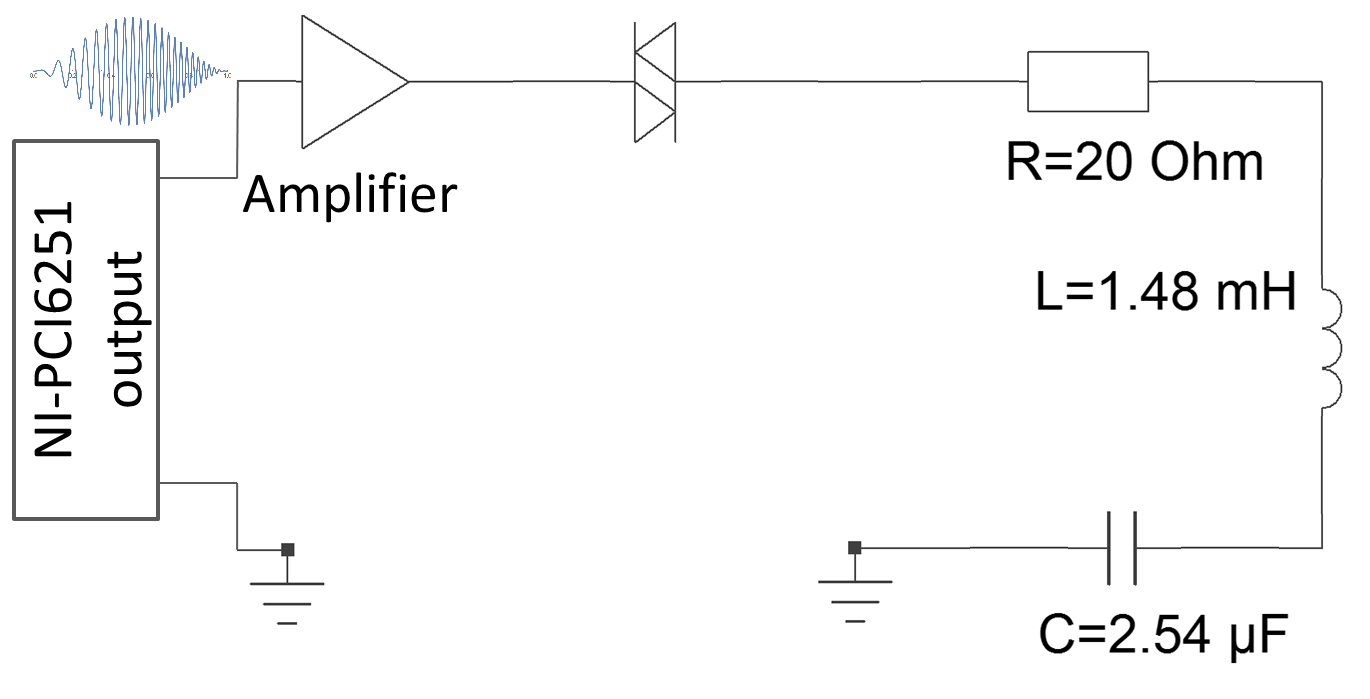}
	\caption{Scheme of the spin-flip electronic device. 
	It includes a NI-PCI6251 Multifunction Acquisition Card, an amplifier, a pair of diodes, a resistor, the spin-flip coil and a capacitor.}
	\label{fig:Signal and scheme}
\end{figure}

Applying spin flips with a transverse oscillating magnetic field to reverse the polarization, one can remove all magnetic contributions which are independent of the gas polarization, such as fluxgates offsets  or a misalignment of the magnetometers' axes with $B_0$.
This well-known technique, called "adiabatic fast passage" \citep{Luschikov1984}, requires a spin-flip coil and a RC circuit resonating at the Larmor frequency, corresponding to a $80~\rm{\mu T}$ holding field.
An extension of this method to spin-flip polarized $^3$He confined in a cell may be found in Ref. \citep{Babcock2007}.
Our spin-flip coil is a solenoid of $50~\rm{cm}$ long and $25~\rm{cm}$ diameter with 50 turns of copper wires.
Eleven additional turns at each end allow us to have a more homogeneous magnetic field in the spin-flip coil.
The total resistance of the device is $21.8~\rm{Ohms}$.
The spin-flip signal we used is an oscillating magnetic field $b_{SF}$ directed orthogonally to the main field $B_0$. 
During the spin-flip signal, the frequency of this signal is swept through the Larmor frequency $f _0=\gamma B_0/2\pi\approx 2.6~\rm{kHz}$ and the signal envelop is modulated using a fourth-order polynomial shape:
\begin{equation}
B_1 (t) = B_{1,\mathrm{max}} \left( \frac{2t}{\tau_{\mathrm{SF}}} \right) ^2 \left( 2- \frac{2t}{\tau_{\mathrm{SF}}} \right) ^2 ,
\label{eq:formePulseSF}
\end{equation}
with $\tau _{\rm{SF}}$ the duration of the spin flip (in our case, $\tau_{\rm{SF}} = 100~\rm{ms}$).
The frequency-sweep range is equal to the spin Larmor frequency. 
This signal is generated by an NI-PCI6251 Multifunction Acquisition Card with an amplitude of $5.5~\rm{V}$.
A voltage amplifier with a gain of 1.86 generates the oscillating current in the resonating circuit. 
To avoid maser effects which correspond to a strong coupling between the highly polarized gas and the spin-flip circuit \citep{Bloom1957}, we used a rather low quality factor ($Q=0.989$) and inserted in the resonating circuit a pair of diodes with low threshold ($\approx 0.4~\rm{V}$) which decouples the spin-flip coil from the rest of the circuit between the spin-flip signals.
Figure \ref{fig:Signal and scheme} presents a scheme of the spin-flip electronic device.

\begin{figure}
\includegraphics[width=0.470\textwidth]{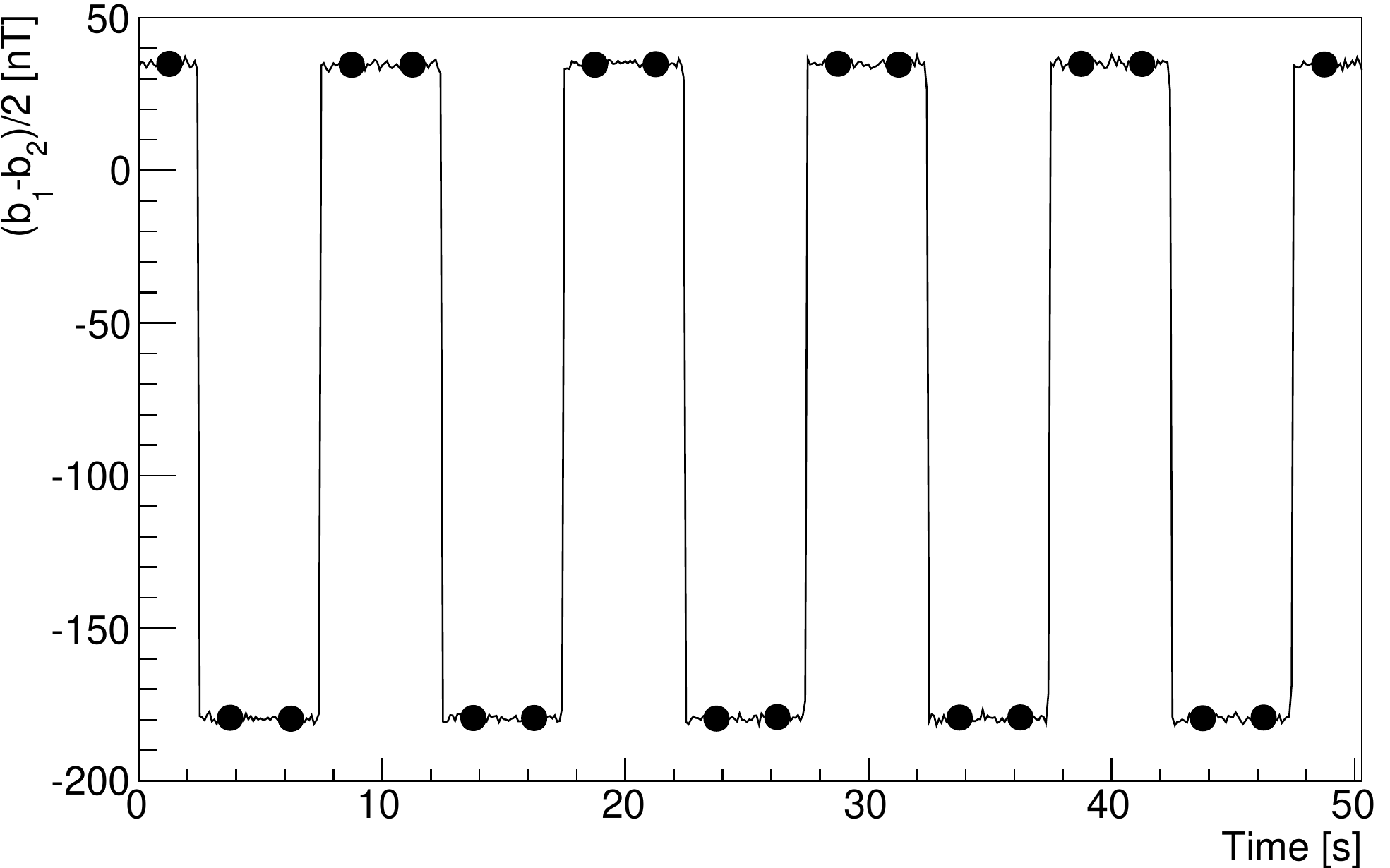}
\caption{\label{fig:pedago}
Typical sequence of measurement of the transverse magnetic field generated by a cylindrical cell at 4 bars with two fluxgate magnetometers.
The upper (lower) points correspond to the spin-up (-down) state of the gas.}
\end{figure}

A polarization measurement sequence consists therefore of measuring the transverse magnetic field $b(\uparrow )$ with the two fluxgates during $1~\rm{s}$, applying a spin flip, measuring twice the magnetic field $b(\downarrow )$, applying a second spin flip and measuring the magnetic field $b(\uparrow )$ ($+--+$ \textit{sequence}).
Figure \ref{fig:pedago} shows typical sequences of $+--+$ measurements of the magnetic field induced by a spherical cell at 1 bar with the two magnetometers.
The difference between the four measurements allows us to access directly to the magnetic field generated by the cell gas.
Since this magnetic field is proportional to the gas polarization, one can deduce the polarization value at any time if given the initial polarization value.
The polarization is expressed in $\mathrm{nT}$; for example, a $5~\rm{cm}$ radius spherical cell filled with $1~\rm{bar}$ $70\%$-polarized gas cell generates a $30~\rm{nT}$ magnetic field which is measured by the magnetometers at the glass surface.
The statistical precision of one polarization measurement sequence, described above, is $0.16~\rm{nT}$.
Repeating eight times this sequence will give a $60~\rm{pT}$ precision and thus a signal to noise ratio of about 1000.

To determine the loss of polarization induced by a spin flip, we measured the loss of polarization induced by a large number of spin flips (about $10^4$) during a short period ($1~\rm{h}$).
We conclude that each spin flip induces a loss of polarization of $3\times 10^{-6}$.
No dependence of these losses with the pressure or the polarization of the gas has been observed.

\subsection{Structure of a measurement run}

\begin{figure}
	\centering
		\includegraphics[width=0.470\textwidth]{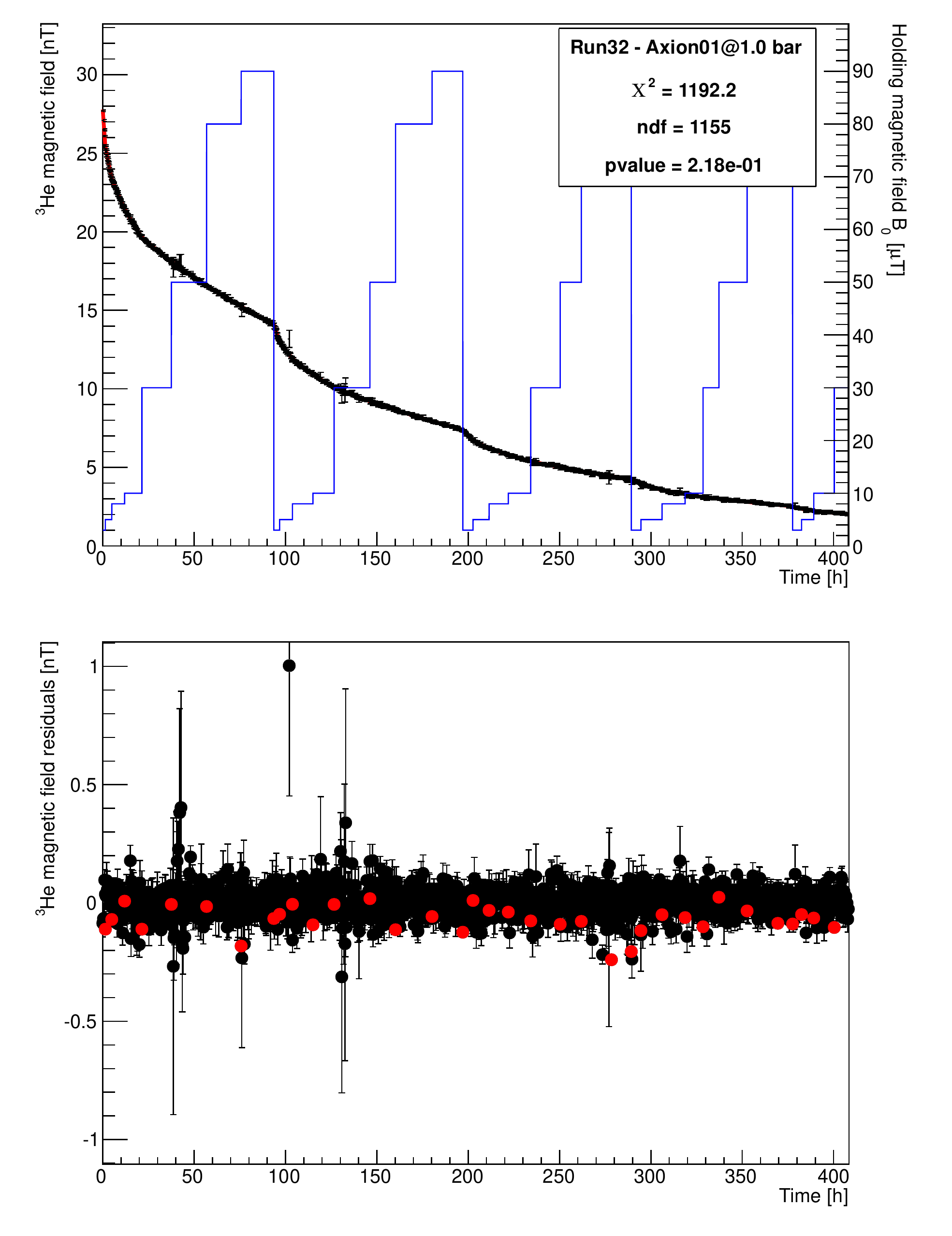}
	\caption{Result of the $\chi ^2$ minimization method for Run 32. 
	Up: the raw data (in black) and the best fit (in red) as a function of time and the holding field. 
	The blue curve corresponds to the holding magnetic field value as a function of time.
	Down: residuals between the data and the polarization reconstruction. 
	The red points correspond to the last measurement point of each $T_1$ measurement.}
	\label{fig:FinalPlot-Run32}
\end{figure}

In the experiment, we measure the relaxation rate of spin-polarized $^3$He gas as a function of the strength of the applied holding magnetic field $B_0$.
In a given holding field value, the polarization measurement is done every $20~\rm{min}$.
When the polarization relative loss is sufficient to precisely estimate the relaxation rate ($\approx 10~\%$), a new cycle begins, and the holding field value is changed.
This procedure is repeated for holding fields between $3$ and $90~\rm{\mu T}$: eight cycles correspond to a set of measurement.
These sets are repeated until the polarization reaches a few $\rm{nT}$.
The combination of all the measured sets of one cell is called \emph{a run}.
The upper plot on Figure \ref{fig:FinalPlot-Run32} gives the result of the Run 32 for a spherical cell, named "Axion01" of $6~\rm{cm}$ radius filled with $1~\rm{bar}$ of $70\%$-polarized $^3$He.
This cell possesses  $3~\rm{mm}$ thick aluminosilicate walls with a caesium coating and a cylindrical appendix, used to fill the cell with the helium gas.
The typical relaxation time induced by the walls of this cell is about $400~\rm{h}$ (see Sec. \ref{sec:wallcollision}).
Because of this long $T_1$, a run using this cell lasts typically between 2 and 3 weeks.
A smaller cell named CCT12 was also used for this experiment.

\section{Analysis and results}

Runs have been performed for different spherical cells at different pressures.
These characteristics are presented in Table \ref{tab:liste-Runs}.
\begin{table}[t]
	\centering
\caption{Main characteristics of the nine Runs obtained with the apparatus.
	The value of the indicated initial polarization was measured on the Tyrex installation using an optical method \citep{Andersen2005}.}
	\label{tab:liste-Runs}
		\begin{tabular}{c|c|c|c|c|c}
		Run			& Cell        & Radius& Pressure & Coating  &  Initial            \\
		        &             & (cm)  & (bar)    &          &  polarization       \\ \hline
		32      & Axion01     & 6		  & 1 	     & Caesium  & $70\%$							\\
		33      & Axion01     & 6		  & 4 	     & Caesium	& $70\%$							\\
		34      & CCT12       & 4	    & 4 	     & Rubidium & $70\%$						  \\
		35      & CCT12       & 4		  & 1 	     & Rubidium & $70\%$						  \\
		36      & Axion01     & 6		  & 2 	     & Rubidium & $70\%$						  \\
		37      & Axion01     & 6		  & 3 	     & Rubidium & $70\%$					    \\	
		38      & BufferAspec & 6	    & 1 	     & None     & $70\%$					  \\	
		39      & Axion01     & 6		  & 0.3 	   & Rubidium & $70\%$						  \\
		41      & Axion01     & 6		  & 3		 	   & Rubidium & $70\%$						  
		\end{tabular}
\end{table}
For each run, the polarization as a function of time and holding field value is measured.
Next, the gas relaxation is studied in order to find a deviation from the expected behavior \eqref{eq:Gamma1-explicite}.

In this section, the reconstruction method of the polarization as a function of time is presented.
The extraction of the relaxation parameters for every run and the constraint on scalar-pseudoscalar coupling constants are then discussed in details.

\subsection{Analysis procedure}

At any time, for a given polarization value and magnetic field, the expected standard relaxation rate $\Gamma _1 (P,B_0)$ is given by Eq. \eqref{eq:Gamma1-explicite}.
If a new interaction  between two nucleons mediated by a pseudoscalar boson exists, a new depolarization channel \eqref{eq:Gamma1NF-formulefinale} will add to the standard one.
Therefore the analysis procedure consists in reconstructing the polarization evolution as a function of time and holding field for a given boson mass or interaction range and compare it with the data.
In the case of a polarization evolving slowly compared with the time between two polarization measurements, this step-by-step reconstruction is given by
\begin{equation}
P^{\rm{mod}}_{i+1} = P^{\rm{mod}}_i \exp \left(-\Gamma_1(P^{\rm{mod}}_i,B_i)\times (t_{i+1}-t_i) \right) ,
\label{eq:construction-polarisation}
\end{equation}
where $P^{\mathrm{mod}}_i$ denotes to the polarization at the time $t_i$.

First, only standard contributions to the relaxation rate \eqref{eq:Gamma1-explicite} are taken into account.
For a given set of parameters $\left\{ a, \cdots , f, P_0\right\}$ with $P_0$ the initial polarization value (expressed in $\mathrm{nT}$), the polarization reconstruction can be compared with the obtained measurements, using a $\chi ^2$  method.
The minimization of such a quantity gives the most likely set of parameters. 
The result of this method is shown on Figure \ref{fig:FinalPlot-Run32}.
For each run, this reconstruction and analysis procedure is applied.
Since the reduced $\chi ^2$ is close to 1, we can conclude no deviation in the data is observed: the standard contributions given by \eqref{eq:Gamma1-explicite} explain the observed evolution of the polarization of each run.

\subsection{Behavior of the fit parameters\label{sec:bahviorfitparam}}

For every run presented in Table \ref{tab:liste-Runs}, we extracted the parameters of the standard depolarization \eqref{eq:Gamma1-explicite} in the Axion01 cell for the different values of pressure.
Among them, the parameters $a$, $c$ and $f$ have explicit physical meanings.
Let us present their behavior with respect to the pressure.

\subsubsection{Atomic and walls collisions depolarization\label{sec:wallcollision}}

For a given pressure in a given cell, the $a$ parameter in \eqref{eq:Gamma1-explicite} corresponds to the sum of atomic $\Gamma _{dd}$ and walls $\Gamma _{w}$ collisions relaxations, the spin-flip induced relaxation and the magnetic depolarization generated by the solenoid gradients.
The first contribution is proportional to the pressure.
The depolarization induced by the spin flips (typically $6\times 10^{-6}~\rm{h^{-1}}$) is independent of the gas polarization and the holding field value and contributes weakly to the parameter $a$ compared with the atomic and walls collisions depolarization.
The last contribution is inversely proportional to the pressure but it is also expected to be very small compared with the two first ones.

\begin{figure}
	\centering
		\includegraphics[width=0.47\textwidth]{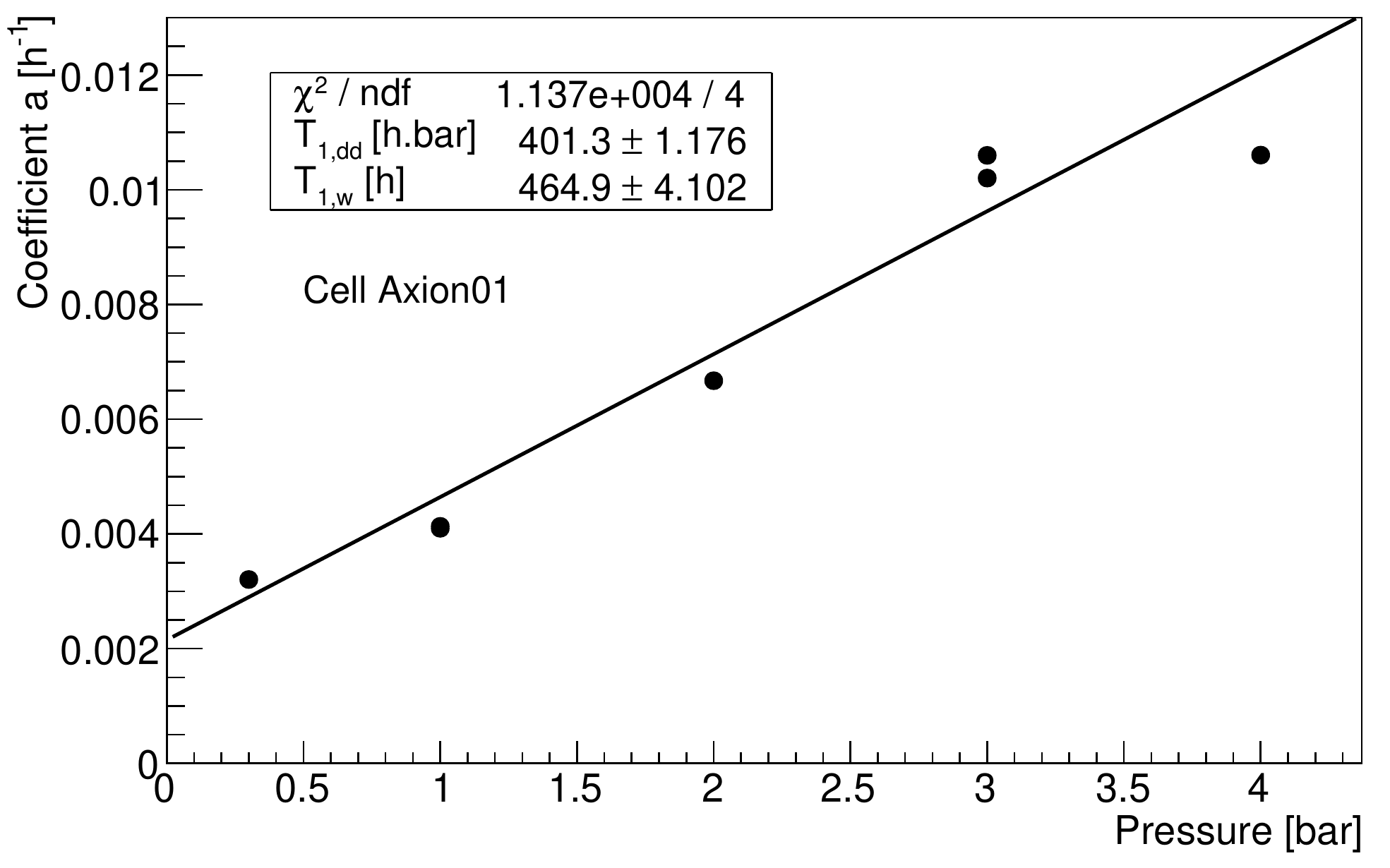}
		\caption{Parameter $a$ in \eqref{eq:Gamma1-explicite} as a function of the pressure in the Axion01 cell. }
	\label{fig:a_vs_Pressure}
\end{figure}

Figure \ref{fig:a_vs_Pressure} presents the  parameters $a$ obtained in Runs 32, 33, 36, 37 and 39 as a function of the pressure in the Axion01 cell.
A linear fit was performed in order to extract the relaxation times induced by the atomic and walls collisions.
The measured atomic collisions relaxation time  is $400~\rm{h}$ for $1~\rm{bar}$ of $^3$He gas .
A theoretical calculation of this contribution for gaseous $^3$He predicts a lifetime of $798~\rm{hour\cdot bar}$ \citep{Newbury1993}.
This value does not correspond to the one extracted from our measurements.
However, the presence of a few ppm of contaminants such as oxygen in the helium gas can explain this difference.
Indeed, collisions between oxygen and helium can cause a new source of polarization relaxation with the same dependence on the pressure as He-He collisions depolarization.

Notice that the statistical error due to the measurement procedure is smaller than the point size on Figure \ref{fig:a_vs_Pressure}, while the data points present a significant spread. 
This can be understood taking into account that each measurement at a new pressure value required a new run of the Tyrex filling station to prepare a new filling with polarized gas. 
This complex procedure cannot guarantee the same $T_1$ due to the unavoidable contamination during the gas compression from $~\rm{mbar}$ (the pressure at which the gas is optically polarized in Tyrex) to the desired pressure in a few bar range.

The walls induce a depolarization time of $460~\rm{h}$ for the Axion01 cell.
This long time is due to the high quality of the rubidium coating on the cell surface; an improvement of the walls quality by a factor of 2 will no longer make walls collisions relaxation a limiting contribution.

\subsubsection{Environment magnetic gradients depolarization}

\begin{figure}
	\centering
		\includegraphics[width=0.47\textwidth]{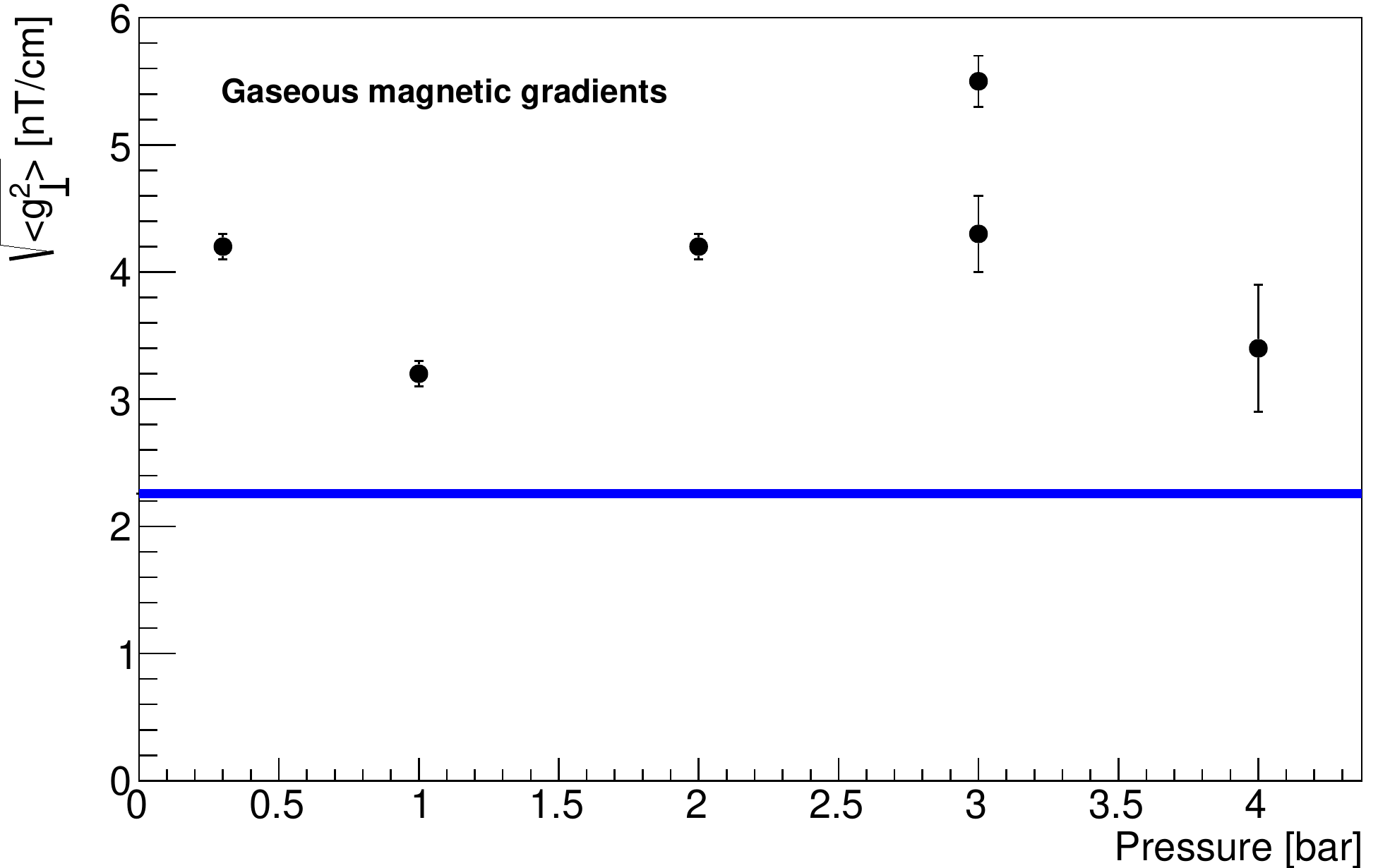}
	\caption{Magnetic fields gradients extracted from the runs involving the Axion01 cell as a function of the pressure, corresponding to the parameter $c$ in \eqref{eq:Gamma1-explicite}.
	The blue line corresponds to the gradient induced by the magnetic environment extracted from the magnetic maps.}
	\label{fig:canSolenoid}
\end{figure}

The third term $c/B_0^2$ in \eqref{eq:Gamma1-explicite} corresponds to relaxation induced by the environmental magnetic inhomogeneities $\vec{b}_{\rm{ext}}$.
Figure \ref{fig:canSolenoid} presents the transverse gradients $\sqrt{\langle g_{\perp}^2 \rangle }$ extracted from $c$ using \eqref{eq:adiabdiff} as a function of the pressure for the Axion01 cell.
The relative external magnetic gradients are typically smaller than $10^{-3}~\rm{cm^{-1}}$, which corresponds to a relaxation time of $100~\rm{h}$ at $3~\rm{\mu T}$.
The difference of a few $\rm{nT/cm}$ between the mapping results presented in Table \ref{tab:RésultatsDesCartographiesDeChampsMagnétiques} and this gradients extraction is mainly due to the addition of material surrounding the cell, such as the spin-flip coil support.

\subsubsection{Gaseous magnetic gradients depolarization}

\begin{figure}
	\centering
		\includegraphics[width=0.47\textwidth]{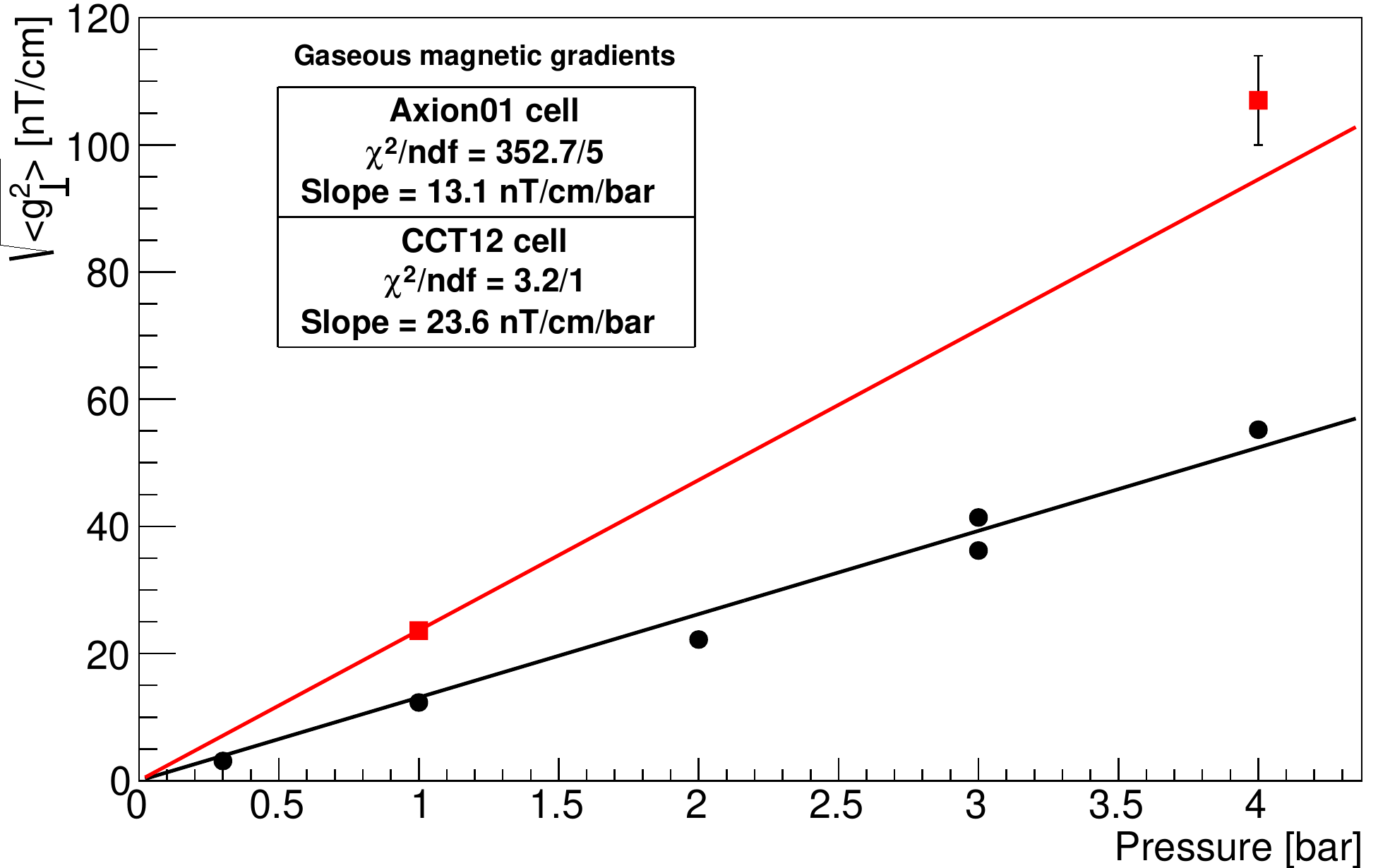}
	\caption{Gaseous magnetic fields gradients extracted from the runs involving the Axion01 cell (black circles) and the CCT12 cell (red squares) as a function of the pressure, corresponding to the parameter $f$ in \eqref{eq:Gamma1-explicite}.}
	\label{fig:canGasMagnetization}
\end{figure}

The relaxation rate \eqref{eq:Gamma1-explicite} contains contributions which depend on the gas polarization $P$.
The last term in this equation corresponds to the depolarization induced by the cell gradients.
Figure \ref{fig:canGasMagnetization} presents the equivalent transverse gradients $\sqrt{\langle g_{\perp}^2 \rangle }$ generated by a $100\%$-polarized gas using \eqref{eq:adiabdiff} as a function of the pressure in the cell.
The magnetic gradients generated by $1~\rm{bar}$ gas inside the Axion01 cell are about $13~\rm{nT/cm}$.
These gradients limit the sensitivity of our experiment at low magnetic field and high polarization.
In order to improve the gradients generated by the gas polarization and improve the sensitivity of the method at high polarization, a dedicated study has been initiated.

\subsection{Extraction of constraints on a new interaction}

\begin{figure}
	\centering
		\includegraphics[width=0.5\textwidth]{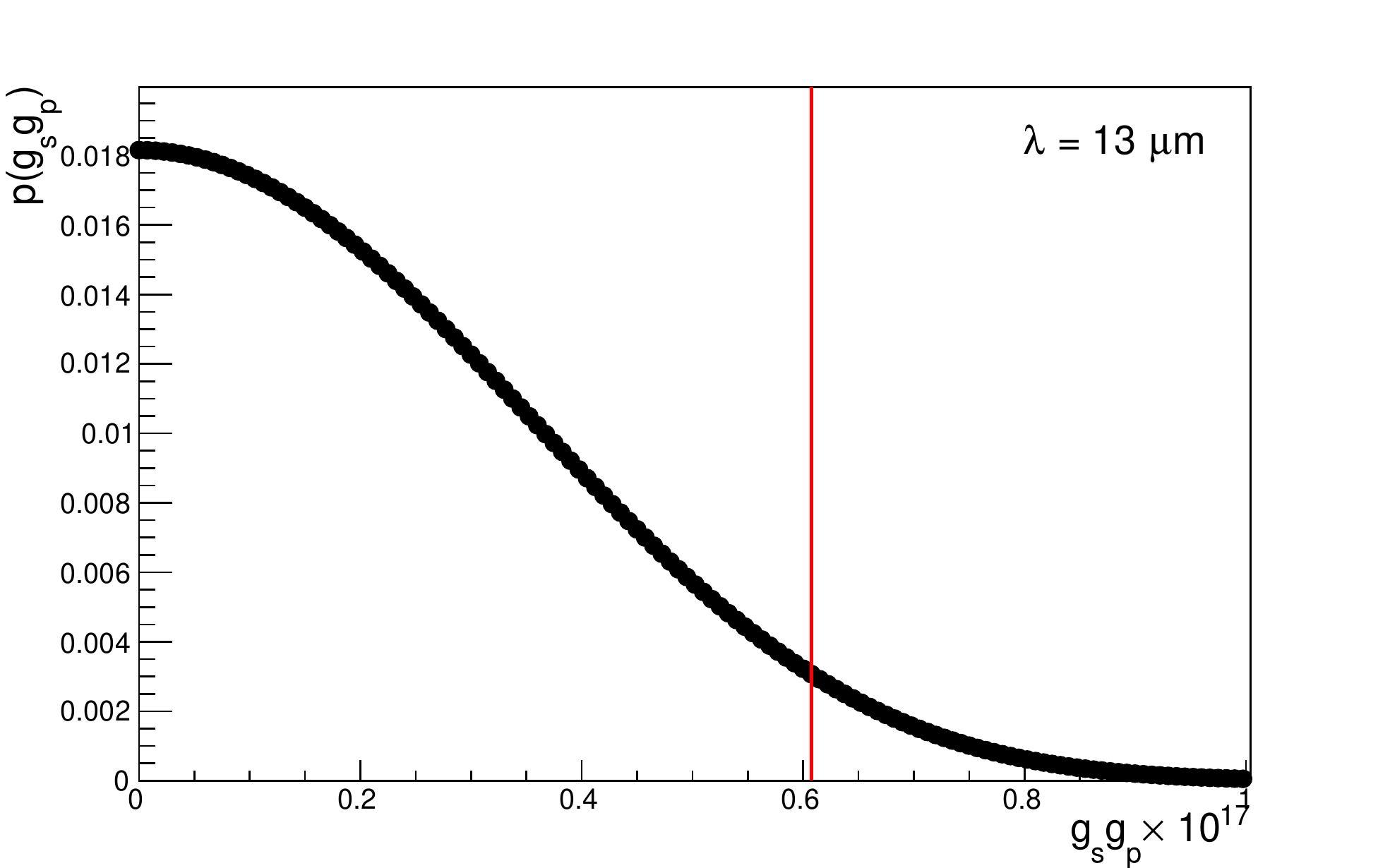}
	\caption{\emph{A posteriori} density function of the $g_sg_p$ parameter for $\lambda = 13~\rm{\mu m}$ obtained from Run 32. The integral to the left to the vertical red line corresponds to a $95\%$ confidence level.}
	\label{fig:Integral-profLikelihood-gsgp-13mum}
\end{figure}

To confirm the absence of a new short-range spin-dependent force which is derived from \eqref{eq:potential}, a Bayesian approach is employed to extract the \emph{a posteriori} probability density function of the parameters $\lambda$ and $g_sg_p$.
The relaxation rate induced by a short-range interaction is added to the standard depolarization contributions.
The likelihood function $\mathcal{L} (\mathrm{data}\vert a, \cdots , f, P_0, \lambda, g_sg_p )$  is built, assuming a Gaussian distribution of the data around the reconstruction model.
Considering flat priors, the probability density function $p(\lambda ,g_sg_p\vert \mathrm{data})$ is obtained by marginalizing the likelihood function over the nuisance parameters $a, \cdots , f$ and $P_0$ and normalizing it to 1.
For each range value $\lambda$ between $1$ and $100~\rm{\mu m}$, the \emph{a posteriori}function is constructed as represented on Figure \ref{fig:Integral-profLikelihood-gsgp-13mum}.
The maximum of this function is for $g_sg_p = 0$, which confirms that there is no evidence of a short-range interaction for the considered range of $\lambda$.
To extract an upper limit on $g_sg_p$ for each $\lambda$ with $95\%$ C.L., an integration of the density function is performed.
The upper limit $g_sg_{p,\mathrm{lim}}$ is defined by
\begin{equation}
\int _0 ^{g_sg_{p,\mathrm{lim}}} p(\lambda, g_sg_p\vert \mathrm{data}) \mathrm{d}g_sg_p = 95\%.
\end{equation}
Using Eq. \eqref{eq:GammaNF_simple} and the data obtained with Run 32, we put a constraint on the product $g_sg_p\lambda ^2$ for ranges $\lambda$ between $1~\mu \mathrm{m}$ and $100~\mu \mathrm{m}$:
\begin{equation}
g_sg_p \lambda ^2 \leq 1.1\times 10^{-27}\,\mathrm{m}^2 ( 95\%\,  \mathrm{C.L.}) .
\end{equation}

\begin{figure}
	\centering
		\includegraphics[width=0.470\textwidth]{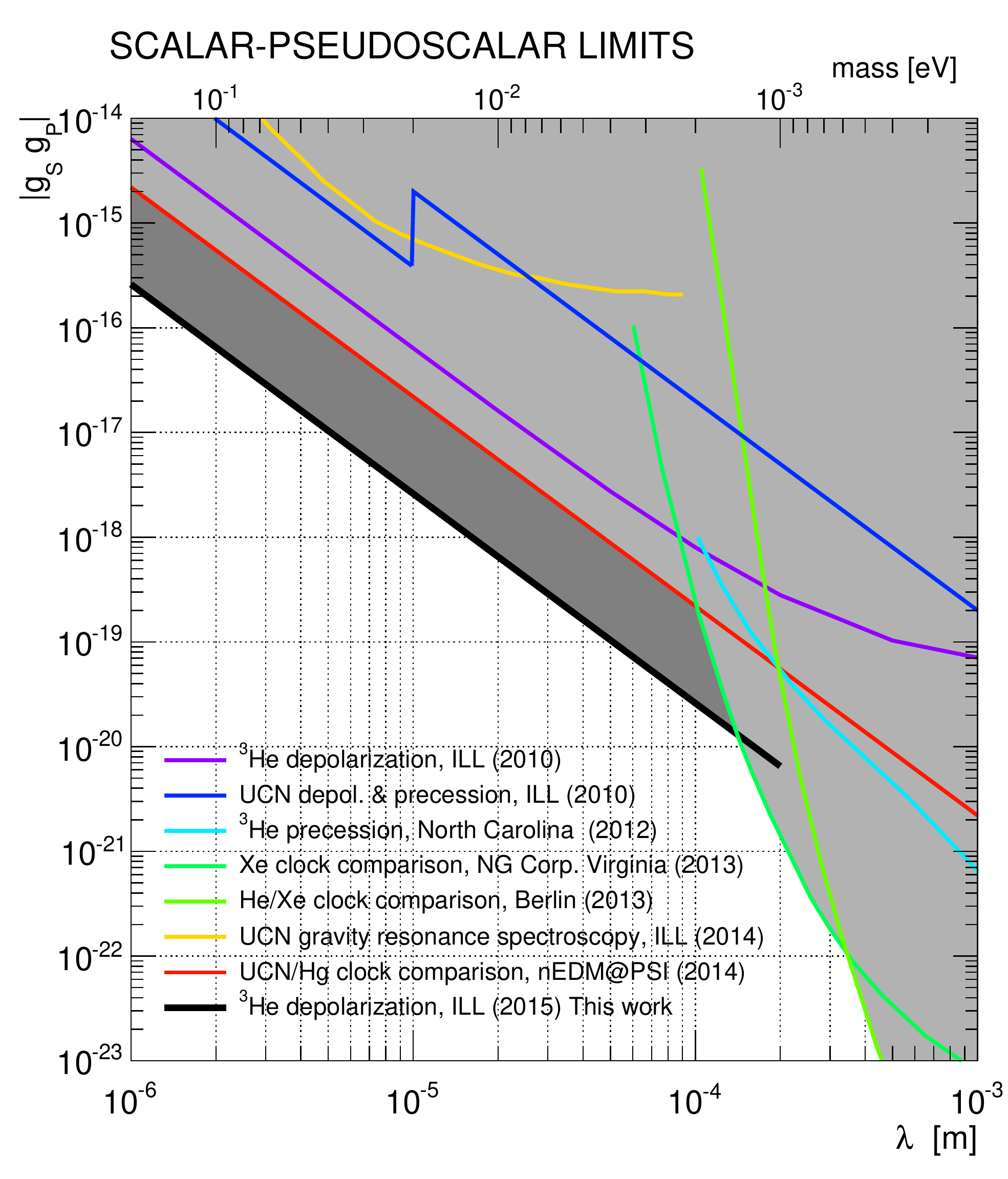}
	\caption{Constraints on the coupling $g_sg_p$ of an interaction mediated by the exchange of a scalar boson as a function of the range $\lambda$ and the mass given by $m_0=\frac{\hbar}{\lambda c}$. 
	Black: this work.
	Purple: $^3$He relaxation rate measurements \citep{Petukhov2010}.
	Dark blue: Ultra-Cold Neutrons (UCN) depolarization and precession \citep{Serebrov2009a}.
	Light blue: $^3$He precession \cite{Chu2012}.
	Light green: clock comparison with Xe isotopes \citep{Bulatowicz2013}.
	Dark green: clock comparison with He and Xe \cite{Tullney2013}
	Yellow: gravitational levels of UCN \citep{Jenke2014}.
	Red: Hg/UCN precession frequencies comparison \citep{Afach2015}.}
	\label{fig:scalarPseudoscalar}
\end{figure}

For each run presented in Table \ref{tab:liste-Runs}, this analysis procedure is applied and no evidence of an exotic depolarization channel has been observed.
Table \ref{tab:liste-RunsConstraints} presents the constraints obtained on $g_sg_p\lambda ^2$ for each run.
\begin{table}[t]
	\centering
	\caption{Constraints on $g_sg_p$ obtained with the nine runs. 
	These constraints are valid for ranges between $1~\rm{\mu m}$ and $100~\rm{\mu m}$.}
	\label{tab:liste-RunsConstraints}
		\begin{tabular}{c|c|c}
		Run			& Cell                & $g_sg_p\lambda$ ($\mathrm{m}^2)$ (95 \% C.L.) \\ \hline
		32      & Axion 01 @ 1 bar   &	$11\times 10^{-28}$					\\
		33      & Axion 01 @ 4 bars   &	$6.7\times 10^{-28}$				\\
		34      & CCT12 @ 4 bars      &	$21\times 10^{-28}$				  \\
		35      & CCT12 @1 bar        &	$11\times 10^{-28}$				      \\
		36      & Axion 01 @ 2 bars   &	$11\times 10^{-28}$						\\
		37      & Axion 01 @ 3 bars   &	$5.0\times 10^{-28}$						\\
		38      & BufferAspec @ 1 bar & $60\times 10^{-28}$						\\
		39      & Axion 01 @ 0.3 bars  &	$11\times 10^{-28}$						\\
		41      & Axion 01 @ 3 bars    &	$4.4\times 10^{-28}$						
		\end{tabular}
\end{table}
A combination of these limits allows us to set a better constraint on the $g_sg_p$ product
\begin{equation}
g_sg_p \lambda ^2 \leq 2.6\times 10^{-28}\,\mathrm{m}^2 ( 95\%\,  \mathrm{C.L.}) .
\end{equation}
Figure \ref{fig:scalarPseudoscalar} shows this constraint for ranges between $1$ and $100~\rm{\mu m}$.

\section{Conclusion}

Measuring $^3$He hyperpolarized gas relaxation as a function of the holding field is a sensitive method to search for short-range spin-dependent exotic interactions.
To fully explore the potential of such a technique, a dedicated experimental setup was built with a particular effort on the reduction of magnetic inhomogeneities and the improvement of the polarization measurement.
A more stringent constraint on the scalar-pseudoscalar coupling of an exotic boson to nucleons has been extracted from these measurements.
Compared with the 2010 experiment \citep{Petukhov2010}, a factor 20 of improvement in terms of sensitivity has been obtained.

A significant improvement of more than 1 order of magnitude in sensitivity on $g_sg_p$ (corresponding to 2 orders of magnitude in terms of relaxation rate measurement) seems quite difficult to achieve, even in a better magnetic shield.
Since our sensitivity is limited by the magnetic gradients generated by the polarized gas, significant efforts must be done in understanding the origin of the gas self-relaxation in order to overcome this limiting obstacle.

 The obtained limit is better by a factor 8 than the previous best constraint obtained by the nEDM apparatus \citep{Afach2015} and a factor 20 compared with the previous experiment performed in 2010 at ILL measuring $^3$He longitudinal relaxation \citep{Petukhov2010}.


\section*{Acknowledgments}
We are thankful to Pierre Fayet, Pierre-Jean Nacher and Genevi\`{e}ve Tastevin for fruitful discussions.
We thank also the "Service D\'etecteurs et Instumentation" (SDI) from LPSC and Helium-3 group from ILL, especially R\'emi Faure, Pascal Mouveau and Antonello Rizo, for technical support.
Part of this work was supported by the AGIR program of Universit\'{e} Joseph Fourier (Grenoble).

\bibliography{article}

\end{document}